\documentclass [12pt]{article}
\usepackage{graphicx,amssymb,amsmath,bm}
\usepackage[symbol*]{footmisc}
\usepackage[cp1251]{inputenc}
\usepackage[english]{babel}
\usepackage{booktabs}

\makeatletter
\def\@biblabel#1{#1.\hskip-0.3em}
\makeatother

\begin{document}
\def\refname{\normalsize \centering \mdseries \bf References}
\def\abstractname{Abstract}

\begin{center}
{\large \bf Test of the Glauber Formula for Nucleon-Deuteron Scattering
at Intermediate Energies}
\end{center}

\begin{center}
\bf\text{V.~I.~Kovalchuk}
\end{center}

\begin{center}
\small
\textit{Department of Physics, Taras Shevchenko National University, Kiev 01033, Ukraine}
\end{center}

\begin{abstract}
A high-precision test of the Glauber formula for the amplitude of nucleon-deuteron
scattering is performed. Nucleon-nucleon amplitudes used in the calculations depend
on the spins of interacting particles, phase shifts, and mixing parameters. These
amplitudes were derived by using the Nijmegen potentials. The differential cross
sections for nucleon-deuteron scattering were calculated for the projectile-nucleon
energies of 65, 95, 135, 150, 190, and 250~MeV, and the results of these calculations
were compared with experimental data.
\end{abstract}

\bigskip
\begin{center}
\bf{1.~Introduction}
\end{center}
\smallskip

Investigation of collisions between high-energy particles and nuclei is an important
problem within which one can study nucleon-density distributions, the dynamics of
nucleon-cluster formation in nuclei, the color properties of quark structures,
and so on. From the microscopic point of view, a consistent description of quantities
observed in these reactions is an extremely difficult theoretical problem~\cite{1}.
In view of this, the number of currently existing successful theoretical approaches
is relatively small here, the Glauber method being among them~\cite{2,3}.
In~\cite{1}, it was indicated, with good reason, that, although this important method
has extensively been used in various scattering problems for more than half a century,
only very few studies, surprisingly as it is, have been devoted to rigorously verifying
the accuracy and the applicability range of the Glauber ansatz. It is the opinion of
the present author that, for a first step along these lines, we could address the problem
of a more rigorous derivation of form for the  nucleon-nucleon amplitude in elastic nucleon-deuteron
scattering, which is the simplest nucleon-nucleus reaction. The amplitude obtained in~\cite{4}
from the conditions of invariance of the nucleon-nucleon interaction with respect to spatial
rotations and space and time inversions could be such a form of representation of the
nucleon-nucleon amplitude. This amplitude has the most general form and depends on the spins of
colliding particles, their energy, phase shifts, and mixing parameters. Using it in the
well-known formula of the eikonal approximation for the amplitude of nucleon scattering
on a deuteron~\cite{3}, we can directly test the correctness of the Glauber approach for
the case being considered.

\bigskip
\begin{center}
\bf{2.~Formalism}
\end{center}
\smallskip

The differential cross section for nucleon-deuteron scattering is calculated here by
the formula~\cite{5} (below, use is everywhere made of the c.~m. frame and the system of
units where $\hbar\!=\!c\!=\!1$)
\begin{equation}
\sigma(\theta)\equiv \frac{d\sigma}{d\Omega}=\frac{1}{6}\,Tr\bigl(F_{d}F_{d}^{\dagger}\bigr),
\label{cs}
\end{equation}
\noindent
where $F_{d}$ is the amplitude for nucleon-deuteron scattering. It has the form~\cite{3}
\begin{equation*}
F_{d}({\bf q})=k_{d}(k_{1}^{-1}\,F_{1}({\bf q})+k_{2}^{-1}\,F_{2}({\bf q}))\,G({\bf q}/2)+
\end{equation*}
\begin{equation}
+\frac{ik_{d}}{2\pi k_{1}k_{2}}\int d^{(2)}{\bf g}\,F_{1}({\bf g} + {\bf q}/2)
F_{2}({\bf g} - {\bf q}/2)G({\bf g}),
\label{fd}
\end{equation}

\begin{equation}
G({\bf g})=\int d^{(3)}{\bf r}\,|\phi_{d}({\bf r})|^{2}\exp(i{\bf g}{\bf r}), \quad
{\bf q},{\bf g}\,\bot\,{{\bf k}_{d}},
\label{form}
\end{equation}
\noindent
where ${\bf k}_{d}$ is the deuteron momentum, $k_{1}$ and $k_{2}$ are the momenta of
the deuteron nucleons, $F_{1,2}$ are the amplitudes for projectile-nucleon scattering
on the deuteron nucleons, ${\bf q}$ is the momentum transfer, and $\phi_{d}({\bf r})$
is the ground-state deuteron wave function.

The sum of the first two terms in expression (\ref{fd}) is the scattering amplitude
in the impulse approximation, $F_{d}^{(i)}$ (it corresponds to taking into account
single collisions between the projectile nucleon and the deuteron nucleons). The last
term in expression (\ref{fd}) is the so-called shadowing correction $F_{d}^{(sh)}$);
it corresponds to the contribution of double scattering to the amplitude, $F_{d}$:
\begin{equation}
F_{d}=F_{d}^{(i)}+F_{d}^{(sh)}.
\label{fd_sum}
\end{equation}

With allowance for the charge invariance of the nucleon-nucleon interaction,
the amplitudes $F_{j}$ ($j=1,2$) have the form~\cite{4,6}
\begin{equation}
F_{j}=f_{1j}+f_{2j}({\bf n}{\boldsymbol\sigma}_{j})({\bf n}{\boldsymbol\sigma})+
if_{3j}\,{\bf n}({\boldsymbol\sigma}_{j}+{\boldsymbol\sigma})+
f_{4j}({\bf m}{\boldsymbol\sigma}_{j})({\bf m}{\boldsymbol\sigma})+
f_{5j}({\bf l}{\boldsymbol\sigma}_{j})({\bf l}{\boldsymbol\sigma}),
\label{fj}
\end{equation}
where $f_{sj}$ ($s=1,2,...,5$) are coefficients that depend on the energy of colliding
nucleons, phase shifts, and mixing parameters (see Appendix A); ${\boldsymbol\sigma}$
is the projectile-nucleon spin operator; and ${\bf n}$, ${\bf m}$, and ${\bf l}$ are
three mutually orthogonal unit vectors defined as
\begin{equation*}
{\bf n}=\frac{{\bf k}_{j}\times{\bf k'}_{j}}{|{\bf k}_{j}\times{\bf k'}_{j}|}, \quad
{\bf m}=\frac{{\bf k}_{j}-{\bf k'}_{j}}{|{\bf k}_{j}-{\bf k'}_{j}|}, \quad
{\bf l}=\frac{{\bf k}_{j}+{\bf k'}_{j}}{|{\bf k}_{j}+{\bf k'}_{j}|},
\end{equation*}
\noindent
where ${\bf k}_{j}$(${\bf k'}_{j}$) is the momentum of the incident (scattered)
$j$th nucleon ($k_{j}=k_{d}/2=k$).

Evaluation of the trace in Eq.~(\ref{cs}) with allowance for Eqs.~(\ref{fd}) and (\ref{fj})
leads to the expression
\begin{equation}
Tr\bigl(f_{d}f_{d}^{\dagger}\bigr)=4|G({\bf q}/2)|^2\,S_{11}+
2(\pi k)^{-1}\,G({\bf q}/2)\,S_{12}+(\pi k)^{-2}\,S_{22},
\label{Sp}
\end{equation}
where $S_{11}$, $S_{12}$, and $S_{22}$ are quantities that depend on the coefficients
appearing in the amplitudes $f_{j}$ (see Appendix B). The calculations of the differential
cross sections in (\ref{cs}) were performed for the projectile-nucleon energies of
65, 95, 135, 150, 190, and 250~MeV by using the nucleon-nucleon phase shifts, mixing parameters,
and deuteron wave functions obtained with the \mbox{Nijm~I}, \mbox{Nijm~II}, \mbox{Nijm~93},
and \mbox{Reid~93} potentials~\cite{7}. Here, we disregarded relativistic corrections, since,
in the energy range being considered, their effect on the cross section is insignificant
(see~\cite{1} and references therein).

\bigskip
\begin{center}
\bf{3.~Analysis of calculation results. Conclusions}
\end{center}
\smallskip

An analysis of the calculated cross sections in the figure leads to the following conclusions:

(i) For a given projectile-nucleon energy $E_{N}$, the relative contribution of double scattering,
$F_{d}^{(sh)}$, is virtually independent of the angle $\theta$ and decreases as $E_{N}$ grows.
In the Glauber approximation (the deuteron size exceeds considerably the range of the nucleon-nucleon interaction),
the shadowing correction decreases more slowly in relation to $F_{d}^{(i)}$ the angle $\theta$ grows~\cite{18,19}.

(ii) It is traditionally assumed~\cite{2,20} that the Glauber formula (\ref{fd}) is valid for $q\!\ll\!{k}$ or
$k\!\ll\!{R_{rms}}$, where $R_{rms}$ is the root-mean-square radius of the deuteron. This condition is independent
of energy and yields the following estimate for the angle $\theta$: $\theta\!<\!10^{\circ}$.
In fact, the results of precise calculations performed in the present study for the aforementioned cross sections
with the above realistic potentials show that the Glauber formalism works well even beyond the original assumptions
of the theory: the calculated curves describe experimental data in the angular range $\theta\!<\!30^{\circ}$ and in
the energy range $65\div150$~MeV; as the projectile-nucleon energy increases further, the angular range becomes
broader: $\theta\!<\!70^{\circ}$ (for 190 and 250~MeV).

\vspace{5mm}
\begin{figure}[!h]
\center
\includegraphics [scale=0.70] {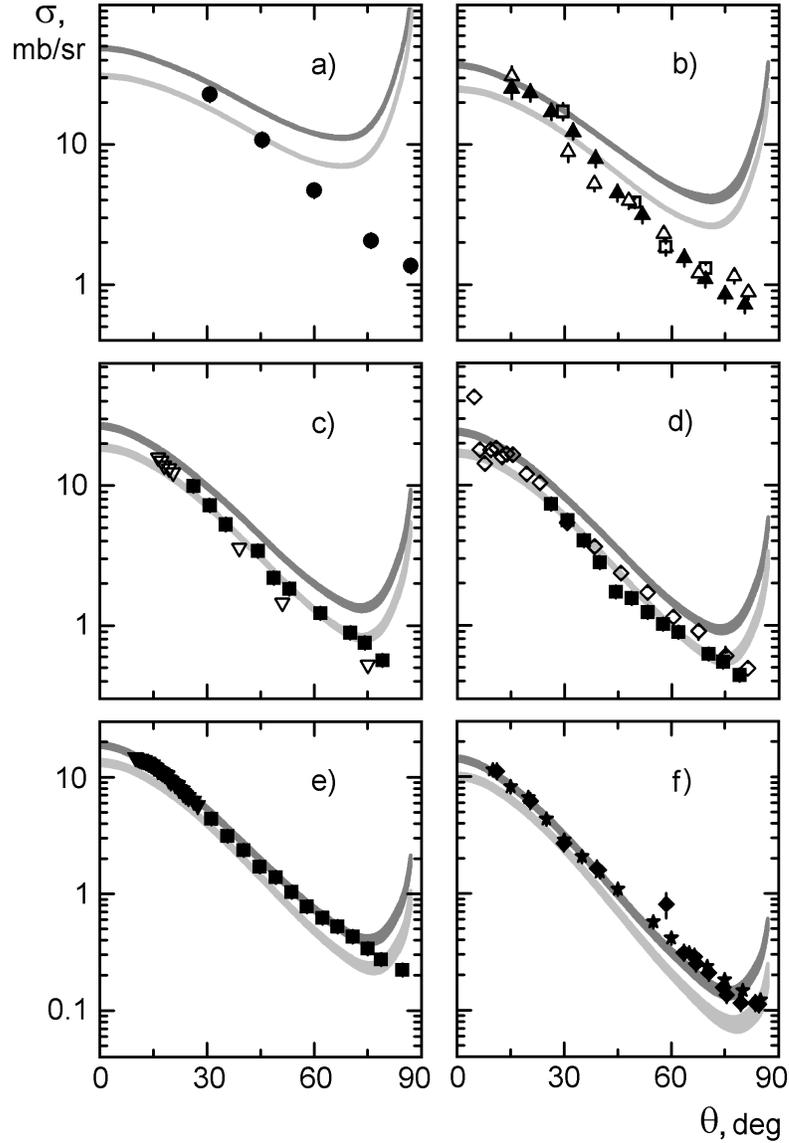}
\caption{Differential cross sections for nucleon-deuteron scattering according to the calculations with
the Nijm~I, Nijm~II, Nijm~93, and Reid~93 potentials for the energies of (a)~65, (b)~95, (c)~135, (d)~150,
(e)~190, and (f)~250 MeV. The cross-section curves lie within the dark-gray bands. The light-gray bands
correspond to the cross sections calculated in the impulse approximation. The displayed experimental data
were borrowed from ($\bullet$)~\cite{8}, ($\blacktriangle$)~\cite{9}, ($\vartriangle$)~\cite{10},
($\square$)~\cite{11}, ($\blacksquare$)~\cite{12}, ($\triangledown$)~\cite{13}, ($\lozenge$)~\cite{14},
($\blacktriangledown$)~\cite{15}, ($\blacklozenge$)~\cite{16}, and ($\bigstar$)~\cite{17}.}
\label{fig1}
\end{figure}

Thus, the test performed here for the Glauber formula in the case of nucleon-deuteron scattering by using realistic
nucleon-nucleon potentials gives sufficient ground to conclude that, in the range of projectile-nucleon energies
$E_{N}$ that is considered here, the theory in question works well beyond the constraint $q\!\ll\!{k}$. Moreover,
the angular range broadens to $\theta\!<\!70^{\circ}$ as soon as the quantity $k^{-1}$ becomes commensurate with
$R_{rms}$, not only much less than it.

\setcounter{section}{0}
\def\theequation{\Alph{section}.\arabic{equation}}
\def\thesection{\normalsize Appendix \Alph{section}}
\setcounter{equation}{0}
\section{}
\hspace{\parindent}
The coefficients $f_{s}$ in the amplitude for nucleon-nucleon scattering in (\ref{fj}) are expressed in terms of the
phase shifts $^{1}\delta_{\ell}^{J}$ and $^{3}\delta_{\ell}^{J}$ and the mixing parameters $\epsilon^{J}$ as
(the index $j$ on $f_{s}$ is suppressed for the sake of convenience)~\cite{4,5}
\begin{equation*}
f_{1}=\frac{1}{4k}\sum\limits_{\ell=0}^\infty{\big[(\ell+2)a_{\ell}^{\ell+1}+(2\ell+1)a_{\ell}^{\ell}+
(\ell-1)a_{\ell}^{\ell-1}+(\ell+1)b_{\ell}^{\ell+1}}+
\end{equation*}
\begin{equation}
+\ell b_{\ell}^{\ell-1}+(2\ell+1)c_{\ell}^{\ell}\big]P_{\ell}(\cos{\theta});
\label{alpha}
\end{equation}

\begin{equation*}
f_{2}=\frac{1}{4k}\bigg\{\sum\limits_{\ell=0}^\infty{\big[(\ell+1)b_{\ell}^{\ell+1}+\ell b_{\ell}^{\ell-1}-
(2\ell+1)c_{\ell}^{\ell}\big]}P_{\ell}(\cos{\theta})-
\end{equation*}
\begin{equation}
-\sum\limits_{\ell=2}^\infty{\bigg[\frac{1}{\ell+1}\,a_{\ell}^{\ell+1}-\frac{2\ell+1}{\ell(\ell+1)}\,
a_{\ell}^{\ell}+\frac{1}{\ell}\,a_{\ell-1}^{\ell}\bigg]P_{\ell}^{2}(\cos{\theta})}\bigg\};
\label{beta}
\end{equation}

\begin{equation}
f_{3}=\frac{1}{4k}\sum\limits_{\ell=1}^\infty{\bigg[\frac{\ell+2}{\ell+1}\,a_{\ell}^{\ell+1}-
\frac{2\ell+1}{\ell(\ell+1)}\,a_{\ell}^{\ell}-\frac{\ell-1}{\ell}\,a_{\ell}^{\ell-1}+
b_{\ell}^{\ell+1}-b_{\ell}^{\ell-1}\bigg]P_{\ell}^{1}(\cos{\theta})};
\label{gamma}
\end{equation}

\begin{equation*}
f_{4}=\frac{1}{4k\cos{\theta}}\bigg\{\sum\limits_{\ell=0}^\infty{\bigg[\frac{1}{2}\big\{(\ell+2)
a_{\ell}^{\ell+1}+(2\ell+1)a_{\ell}^{\ell}+(\ell-1)a_{\ell}^{\ell-1}\big\}(\cos{\theta}-1)}+
\end{equation*}
\begin{equation*}
+(\ell+1)b_{\ell}^{\ell+1}+\ell b_{\ell}^{\ell-1}-(2\ell+1)c_{\ell}^{\ell}\cos{\theta}\bigg]
P_{\ell}(\cos{\theta})+
\end{equation*}
\begin{equation}
+\frac{1}{2}\sum\limits_{\ell=2}^\infty{\bigg[\frac{1}{\ell+1}\,a_{\ell}^{\ell+1}-
\frac{2\ell+1}{\ell(\ell+1)}\,a_{\ell}^{\ell}+\frac{1}{\ell}\,a_{\ell}^{\ell-1}\bigg]
(1+\cos{\theta})P_{\ell}^{2}(\cos{\theta})\bigg\}};
\label{delta}
\end{equation}

\begin{equation*}
f_{5}=\frac{1}{4k\cos{\theta}}\bigg\{\sum\limits_{\ell=0}^\infty{\bigg[\frac{1}{2}\big\{(\ell+2)
a_{\ell}^{\ell+1}+(2\ell+1)a_{\ell}^{\ell}+(\ell-1)a_{\ell}^{\ell-1}\big\}(1+\cos{\theta})}-
\end{equation*}
\begin{equation*}
-(\ell+1)b_{\ell}^{\ell+1}-\ell b_{\ell}^{\ell-1}-(2\ell+1)c_{\ell}^{\ell}\cos{\theta}\bigg]
P_{\ell}(\cos{\theta})+
\end{equation*}
\begin{equation}
+\frac{1}{2}\sum\limits_{\ell=2}^\infty{\bigg[\frac{1}{\ell+1}\,a_{\ell}^{\ell+1}-
\frac{2\ell+1}{\ell(\ell+1)}\,a_{\ell}^{\ell}+\frac{1}{\ell}\,a_{\ell}^{\ell-1}\bigg]
(\cos{\theta}-1)P_{\ell}^{2}(\cos{\theta})\bigg\}}.
\label{epsilon}
\end{equation}

In expressions (\ref{alpha})-(\ref{epsilon}), we have introduced the following notation:
\begin{equation*}
a_{\ell}^{J=\ell}\equiv\sin{^{3}\delta_{J}^{J}}\exp{(i\,{^{3}\delta_{J}^{J}})},
\end{equation*}
\begin{equation*}
a_{\ell}^{J=\ell+1}\equiv\alpha^{J}\cos^{2}{\epsilon^{J}}+\beta^{J}\sin^{2}{\epsilon^{J}}+
\frac{1}{2}\sqrt{\frac{J}{J+1}}(\alpha^{J}-\beta^{J})\sin{2\epsilon^{J}},
\end{equation*}
\begin{equation*}
b_{\ell}^{J=\ell+1}\equiv\alpha^{J}\cos^{2}{\epsilon^{J}}+\beta^{J}\sin^{2}{\epsilon^{J}}-
\frac{1}{2}\sqrt{\frac{J+1}{J}}(\alpha^{J}-\beta^{J})\sin{2\epsilon^{J}},
\end{equation*}
\begin{equation*}
a_{\ell}^{J=\ell-1}\equiv\alpha^{J}\sin^{2}{\epsilon^{J}}+\beta^{J}\cos^{2}{\epsilon^{J}}+
\frac{1}{2}\sqrt{\frac{J+1}{J}}(\alpha^{J}-\beta^{J})\sin{2\epsilon^{J}},
\end{equation*}
\begin{equation*}
b_{\ell}^{J=\ell-1}\equiv\alpha^{J}\sin^{2}{\epsilon^{J}}+\beta^{J}\cos^{2}{\epsilon^{J}}-
\frac{1}{2}\sqrt{\frac{J}{J+1}}(\alpha^{J}-\beta^{J})\sin{2\epsilon^{J}},
\end{equation*}
\begin{equation*}
\alpha^{J}\equiv\sin{^{3}\delta_{J-1}^{J}}\exp{(i\,{^{3}\delta_{J-1}^{J}})},
\end{equation*}
\begin{equation*}
\beta^{J}\equiv\sin{^{3}\delta_{J+1}^{J}}\exp{(i\,{^{3}\delta_{J+1}^{J}})},
\end{equation*}
\begin{equation*}
c_{\ell}^{J=\ell}\equiv\sin{^{1}\delta_{J}^{J}}\exp{(i\,{^{1}\delta_{J}^{J}})}.
\end{equation*}

\setcounter{equation}{0}
\section{}
\hspace{\parindent}
The quantities $S_{11}$, $S_{12}$, and $S_{22}$ in (\ref{Sp}) have the form
\begin{equation}
S_{11}=4\big\{\Re(\alpha_{1}\bar{\alpha}_{2}+\gamma_{1}\bar{\gamma}_{2})+
\sum\limits_{j=1}^{2}(|\alpha_{j}|^{2}/2+|\beta_{j}|^{2}+|\gamma_{j}|^{2}+
|\delta_{j}|^{2}+|\epsilon_{j}|^{2})\big\};
\label{S11}
\end{equation}
\begin{equation*}
S_{12}=8\Im(\bar{\beta}_{1}a_{21}+\bar{\gamma}_{1}a_{31}+\bar{\delta}_{1}a_{41}+\bar{\epsilon}_{1}a_{51})+
\end{equation*}
\begin{equation*}
+8\Im(\bar{\beta}_{2}a_{12}+\bar{\gamma}_{2}a_{13}+\bar{\delta}_{2}a_{14}+\bar{\epsilon}_{2}a_{15})+
\end{equation*}
\begin{equation}
+4\Im\big\{(\bar{\alpha}_{1}+\bar{\alpha}_{2})(a_{11}+a_{33})\big\};
\label{S12}
\end{equation}
\begin{equation*}
S_{22}=2|a_{11}|^2+4|a_{12}|^2+4|a_{13}|^2+4|a_{14}|^2+4|a_{15}|^2+
\end{equation*}
\begin{equation*}
+4|a_{21}|^2+8|a_{22}|^2+12|a_{23}|^2+16|a_{24}|^2+16|a_{25}|^2+
\end{equation*}
\begin{equation*}
+4|a_{31}|^2+12|a_{32}|^2+14|a_{33}|^2+16|a_{34}|^2+16|a_{35}|^2+
\end{equation*}
\begin{equation*}
+4|a_{41}|^2+16|a_{42}|^2+16|a_{43}|^2+8|a_{44}|^2+16|a_{45}|^2+
\end{equation*}
\begin{equation*}
+4|a_{51}|^2+16|a_{52}|^2+16|a_{53}|^2+16|a_{54}|^2+8|a_{55}|^2+
\end{equation*}
\begin{equation*}
+4\Re[(\bar{a}_{13}+2\bar{a}_{23})(a_{31}+2a_{32}]-
\end{equation*}
\begin{equation}
-4\Re[\bar{a}_{33}(a_{11}+2(a_{12}+a_{21})+4a_{22})],
\label{S22}
\end{equation}
where $a_{ij}$ ($i,j=1,2,...,5$) are double integrals of the form
\begin{equation}
a_{ij}=\int{d^{(2)}{\bf g}\,G({\bf g})\,f_{i1}({\bf g}+{\bf q}/2)
f_{j2}({\bf g}-{\bf q}/2)}.
\label{aij}
\end{equation}

\end{document}